\documentclass{emulateapj}
\shorttitle{Inner gaseous disk around MWC 147}
\shortauthors{Bagnoli et al.}

\begin{document}

\title{An inner gaseous disk around the Herbig Be star MWC~147}

\author{ T. Bagnoli\altaffilmark{1}, R. van Lieshout\altaffilmark{1}, 
L. B. F. M. Waters\altaffilmark{1,2}, G. van der Plas\altaffilmark{1},
B. Acke\altaffilmark{2}, H. van Winckel\altaffilmark{2}, G. Raskin\altaffilmark{2},
P. D. Meerburg\altaffilmark{1,3} }

\affil{$^1$ Astronomical Institute ``Anton Pannekoek'', University of Amsterdam,
Science Park 904, 1098 XH Amsterdam, The Netherlands}

\affil{$^2$ Instituut voor Sterrenkunde, K.U. Leuven, Celestijnenlaan 200 D, 3001 Leuven, Belgium}

\affil{$^3$ Institute for Theoretical Physics, University of Amsterdam,
Valckeniersstraat 65, 1018 XE Amsterdam, The Netherlands}

\begin{abstract}
We present high-spectral-resolution, optical spectra of the Herbig Be star MWC~147,
in which we spectrally resolve several emission lines, including the [\ion{O}{1}] lines at 6300 and 6363~\r{A}.
Their highly symmetric, double-peaked line profiles indicate that the emission originates in a rotating circumstellar disk.
We deconvolve the Doppler-broadened [\ion{O}{1}] emission lines
to obtain a measure of emission as a function of distance from the central star.
The resulting radial surface brightness profiles are in agreement with a disk structure
consisting of a flat, inner, gaseous disk and a flared, outer, dust disk.
The transition between these components at 2 to 3~AU corresponds to the estimated dust sublimation radius.
The width of the double-peaked \ion{Mg}{2} line at 4481~\r{A} suggests
that the inner disk extends to at least 0.10~AU, close to the corotation radius.
\end{abstract}


\section{Introduction}

Herbig Ae/Be (HAeBe) stars are intermediate-mass ($2 - 8~M_{\sun}$), pre-main sequence stars,
known to be surrounded by circumstellar disks of gas and dust where planets may form \citep[see e.g.][]{wat98, nat00}.
The details of the disk structure, especially the innermost region ($R \lesssim 1$~AU), are still a matter of debate.

For about a decade, near-infrared (NIR) interferometers have been able to resolve the disk at $\sim1$~AU scales.
These NIR interferometric observations, together with NIR spectral energy distribution (SED) modeling,
have led to disk models for HAe and late type HBe stars,
featuring an optically-thin inner hole and a ``puffed-up'' inner dust wall at the dust sublimation radius
\citep[and references therein]{mil07}.
The interferometric observations of some early type HBe stars, however,
cannot be explained by this model, and the nature of their inner disks is still unclear.
In a broader context, the HBe stars form a transition between lower mass T Tauri and HAe stars,
and more massive young stellar objects.
Fundamental processes in disk structure and evolution, such as grain growth, disk dissipation, and possibly planet formation,
may change in the mid-B spectral range.
This is probably related to the effects of shorter evolutionary timescales
and stronger stellar radiation fields \citep[e.g.][]{gor09}.
An interesting object in the transitional regime is MWC~147 (alias HD~259431).

MWC~147 was classified as a B6 type star by \citet{her04}.
They estimate its mass at 6.6~$M_{\sun}$ and age at $3.2 \times 10^5$~yr,
assuming a distance of 800~pc and $R_V = 3.1$.
Compared to other HAeBe stars, MWC~147 seems to have a relatively high mass accretion rate
($4.1 \times 10^{-7}~M_{\sun}~\mathrm{yr}^{-1}$ \citealp{bri07};
$7 \times 10^{-6}~M_{\sun}~\mathrm{yr}^{-1}$ \citealp{kra08}).
There is also evidence for a strong stellar wind \citep{bou03},
with an associated mass loss rate of $\sim10^{-7}~M_{\sun}~\mathrm{yr}^{-1}$ \citep{ski93, nis95}.
Several studies \citep{pol02, bou03} find evidence for a flaring circumstellar disk around the object.
Recently, \citet{kra08} proposed a disk structure composed of
an outer, flared, irradiated dust disk and an inner, optically thick, gaseous accretion disk.
They find that the disk should be notably inclined ($i \approx 40\degr - 60\degr$).

In this paper, we present a description and analysis of various emission lines in the optical spectrum of MWC~147,
resolved at very high spectral resolution.
Emission features can be used to probe the gas component of the circumstellar environment of HAeBe stars,
carrying information about its structure and dynamics.
Section \ref{sect:obs} introduces our observations and explains how the data were reduced.
In Section \ref{sect:analys}, we describe the most prominent spectral features
and apply a method developed by \citet{ack05} and \citet{ack06}
to resolve a rotating circumstellar disk from forbidden oxygen lines.
Finally, in Section \ref{sect:discuss}, we discuss our results in the light of previous research and
draw conclusions about the nature of the disk around MWC~147.

\section{Observations and data reduction} \label{sect:obs}

We obtained two echelle spectra of MWC~147
using the HERMES spectrograph \citep{ras08} mounted on the 1.2~m Mercator telescope,
located at La Palma, Spain.
The instrument covers the full optical wavelength range from 3770 to 9000~\r{A}
with a spectral resolution of $\lambda / \Delta\lambda \approx 85000$ in High Resolution Mode.
The observations (45 minutes exposures) were performed on 2009 October 14 and 19,
and the resulting spectra have signal-to-noise ratios
(S/N; determined around 6300~\r{A}) of about 170 and 140 respectively.
The HERMES pipeline was used for initial data reduction, including
order extraction, background subtraction, flat-field correction, and wavelength calibration.

Further data reduction was performed using IRAF.
For most of our analysis we used the average-combined spectrum of both observations
to achieve a higher S/N of about 220.
After determining the continuum using standard IRAF routines,
the spectra were normalized to unity.
Telluric absorption lines and [\ion{O}{1}] emission from Earth's atmosphere were
removed by fitting and subtracting a Gaussian.
Spectral regions polluted by cosmic rays were cut out.

We determined the radial velocity of MWC~147
by comparing the observed and theoretical line centers
of two photospheric \ion{He}{1} absorption lines at 4471 and 5876~\r{A}.
Line centers were measured by fitting Gaussians to the line profiles,
and an error was calculated from the variance of several measurements each using a different continuum estimate.
We find a heliocentric radial velocity of $21.4 \pm 0.3$ km~s$^{-1}$, which is consistent with cataloged values
($19.00 \pm 4.1$ km~s$^{-1}$ \citealp{bar00}; $23 \pm 2$ km~s$^{-1}$ \citealp{vie94}).
The spectra shown in this paper were converted to velocity space
and centered on the photospheric velocity of the star.

\section{Analysis} \label{sect:analys}

\subsection{The optical spectrum of MWC~147}

\begin{figure}[t!]
\plotone{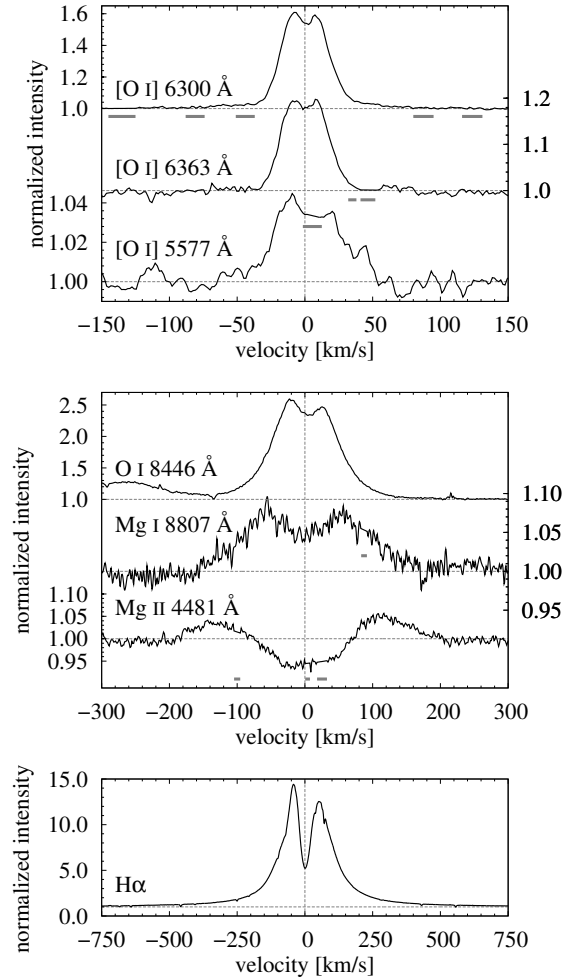}
\caption{Line profiles of several emission lines in the MWC~147 spectrum.
The lines are centered on the stellar radial velocity, indicated by the vertical dotted line.
The continuum is shown by horizontal dotted lines.
The location of the telluric features and cosmic rays that were removed
is indicated by thick horizontal bars.
For the [\ion{O}{1}] 5577~\r{A} line only the observation of 19 October was used,
since the other observation was severely polluted.
The spectrum of the [\ion{O}{1}] 5577~\r{A} line was smoothed using a 5-point boxcar average.
The \ion{O}{1} 8446~\r{A} line is blended with P18 in the blue wing.
Note the different intensity and velocity scales.
\label{fig:lines}}
\end{figure}

\begin{figure}[t!]
\plotone{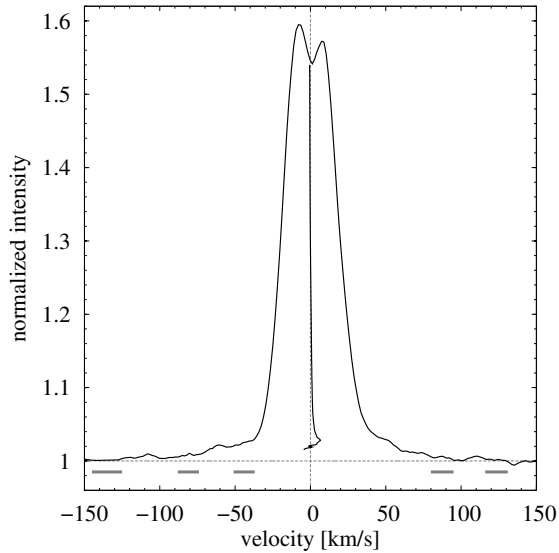}
\caption{The [\ion{O}{1}] 6300~\r{A} line with its bisector.
The spectrum was smoothed using a 5-point boxcar average to suppress noise,
allowing the bisector to be drawn further down.
The horizontal dotted line indicates the continuum level,
the vertical dotted line the stellar radial velocity.
Removed telluric absorption lines and cosmic rays are indicated by thick horizontal bars.
Note the high degree of symmetry of the line.
\label{fig:bisector}}
\end{figure}

The spectrum of MWC~147 contains numerous emission lines, including both permitted
and forbidden transitions of various elements, many of which are double-peaked.
The emission lines most relevant to our analysis ([\ion{O}{1}] 6300, 6363, 5577~\r{A}, \ion{O}{1} 8446~\r{A},
\ion{Mg}{1} 8807~\r{A}, \ion{Mg}{2} 4481~\r{A} and H$\alpha$) are shown in Figure \ref{fig:lines}.
For these lines we determined the equivalent width ($EW$),
full width at half maximum ($FWHM$) and full width at 10\% of maximum intensity ($FW.1M$).
An error on the $EW$s is calculated from the variance of several measurements using different continuum estimates.
The resulting values are presented in Table \ref{tbl:lines}.
For the [\ion{O}{1}] 6300~\r{A} line we furthermore assessed the degree of symmetry by
drawing its bisector (see Figure \ref{fig:bisector}).
Note that the [\ion{O}{1}] 5577~\r{A} line is usually undetected in HAeBe stars,
a relevant point for determining the origin of the [\ion{O}{1}] emission (see Section \ref{origin}).

\begin{deluxetable}{lcccc}
\tablecaption{Quantitative properties of the emission lines shown in Figure \ref{fig:lines}.
\label{tbl:lines}}
\tablewidth{0pt}
\tablehead{
\colhead{Line} & \colhead{$EW$}    & \colhead{$FWHM$}        & \colhead{$FW.1M$}       \\
               & \colhead{[\r{A}]} & \colhead{[km s$^{-1}$]} & \colhead{[km s$^{-1}$]} 
}
\startdata
[\ion{O}{1}] 6300~\r{A}   &  $-0.564\pm0.003$ &  39 &  64 \\ 
$[$\ion{O}{1}] 6363~\r{A} &  $-0.173\pm0.003$ &  39 &  61 \\ 
$[$\ion{O}{1}] 5577~\r{A} &  $-0.044\pm0.001$ &  51 &  89 \\
\ion{O}{1} 8446~\r{A}     &  $-5.300\pm0.047$ & 106 & 201 \\
\ion{Mg}{1} 8807~\r{A}    &  $-0.422\pm0.011$ & 190 & 302 \\
\ion{Mg}{2} 4481~\r{A}    &  $-0.019\pm0.015$ & 312 & 378 \\
H$\alpha$                 & $-63.202\pm0.099$ & 190 & 422 \\
\enddata
\end{deluxetable}

% % This is the table in normal \table markup
% \begin{table}[t!]
% \caption{Quantitative properties of the emission lines shown in Figure \ref{fig:lines}.
% \label{tbl:lines}}
% \centering
% \resizebox{\columnwidth}{!}{%
% \begin{tabular}{lcccc}
% \tableline\tableline
% Line			 & $EW$			 & $FWHM$	 & $FW.1M$	 \\
% 			 & [\r{A}]		 & [km s$^{-1}$] & [km s$^{-1}$] \\ 
% \tableline
% [\ion{O}{1}] 6300~\r{A}	 & $-0.564\pm0.003$	 & 39		 & 64		 \\ 
% $[$\ion{O}{1}] 6363~\r{A}	 & $-0.173\pm0.003$	 & 39		 & 61		 \\ 
% $[$\ion{O}{1}] 5577~\r{A}	 & $-0.044\pm0.001$	 & 51		 & 89		 \\
% \ion{O}{1} 8446~\r{A}	 & $-5.300\pm0.047$	 & 106		 & 201		 \\
% \ion{Mg}{1} 8807~\r{A}	 & $-0.422\pm0.011$	 & 190		 & 302		 \\
% \ion{Mg}{2} 4481~\r{A}	 & $-0.019\pm0.015$	 & 312		 & 378		 \\
% H$\alpha$		 & $-63.202\pm0.099$	 & 190		 & 422
% \end{tabular}}
% \end{table}

The emission lines shown in Figure \ref{fig:lines} all feature double-peaked, extremely symmetric profiles,
except \ion{Mg}{2} 4481~\r{A}, in which the core of the emission is affected by photospheric absorption.
These line profiles strongly suggest the presence of a circumstellar disk in Keplerian rotation.
The [\ion{O}{1}] 6300~\r{A} line consists of two components.
The bulk of its emission is found at low velocities (less than about 35~km~s$^{-1}$),
while weak wings extend up to at least 70~km~s$^{-1}$.
This morphology is examined in more detail in Section \ref{sect:profile}.

If the emission lines emanate from a circumstellar disk, the line width is dominated by rotation
and indicates at what radial distance from the star the emission originates.
In this scenario, the narrow forbidden lines are created furthest away from the star
where gas densities are sufficiently low,
followed by the broader permitted lines of neutral species closer to the star.
The ionized Mg line emanates from the region closest to the star,
consistent with the higher temperature required for the ionization.

Assuming a stellar mass of 6.6~$M_{\sun}$ and a Keplerian rotating disk with an inclination of $50\degr$,
the $FW.1M$ of the \ion{Mg}{2} 4481~\r{A} line corresponds to
an inner radius of the emission region of about 0.10~AU.
This is quite close to the corotation radius
(where the Keplerian rotating disk has the same angular velocity as the rigidly rotating star)
of approximately 0.07~AU,
found using a stellar rotational velocity corresponding to $v \sin i = 90$~km~s$^{-1}$ \citep{boe95},
and a stellar radius of 6.6~$R_{\sun}$, calculated from the luminosity and effective temperature estimates of \citet{her04}.
The H$\alpha$ line may be broadened significantly by electron scattering and should therefore not be analyzed in this way.

\subsection{Origin of [\ion{O}{1}] emission} \label{origin}

\citet{sto00} identified two mechanisms that contribute to
[\ion{O}{1}] emission in photodissociation regions.
In high temperature environments ($T > 3000$~K),
oxygen atoms can be thermally excited due to collisions with hydrogen atoms and free electrons.
Alternatively, the excited oxygen atoms can be the result of
photodissociation of OH molecules by ultraviolet (UV) radiation.

[\ion{O}{1}] emission in HAeBe stars was studied in detail by \citet{ack05}.
Based on the non-detection of the [\ion{O}{1}] 5577~\r{A} line
in all sample stars with coverage at this wavelength,
they derive an upper excitation temperature ($T_{exc}$) limit from the upper 
limit on the 6300~\r{A} to 5577~\r{A} line flux (LF) ratio,
assuming the emission is fully thermal.
Since the resulting $T_{exc}$ limit is lower than 3000~K, they exclude the thermal channel,
and propose that the UV radiation from the star provides all the excited oxygen atoms.

In the case of MWC~147, however, the observed $LF(6300)/LF(5577)$ ratio of about 10
corresponds to a $T_{exc}$ of 4000 to 5000~K,
assuming the [\ion{O}{1}] emission is fully thermal.
Therefore, we cannot exclude that thermally excited atoms contribute to the emission.
The reason for this may be the higher luminosity and temperature of MWC~147.

~

\subsection{The [\ion{O}{1}] surface brightness profile} \label{sect:profile}

We use the [\ion{O}{1}] lines at 6300 and 6363~\r{A} to derive the spatial distribution of the \ion{O}{1} atoms in the disk.
These emission lines are well suited for this purpose, because they probe the (irradiated) disk surface, are optically thin,
and do not suffer from underlying photospheric absorption.
Their spectral shape is a consequence of the velocity field of the emitting region.
By assuming the velocity field to be Keplerian and providing a stellar mass and disk inclination,
it is possible to derive the spatial distribution of [\ion{O}{1}] emission from the line profiles.
First, the red and blue wing of the line are averaged.
Then, the highest-velocity bin that contains [\ion{O}{1}] flux 
is attributed to the innermost ring of the disk, which rotates the fastest.
Due to projection effects, this ring adds emission to the whole line profile.
Its theoretical contribution is computed and removed,
and the procedure is iterated on the residuals of the line from the extrema inwards,
until the whole line profile is dismantled to its very core at $v=0$.
This \textit{spectral deconvolution} method,
based on the model for [\ion{O}{1}] emission in HAeBe stars of \citet{ack05},
is explained in more detail by \citet[their Section 4.2]{ack06}, and applied to three more stars by \citet{vdp08}.

\begin{figure}[t]
\plotone{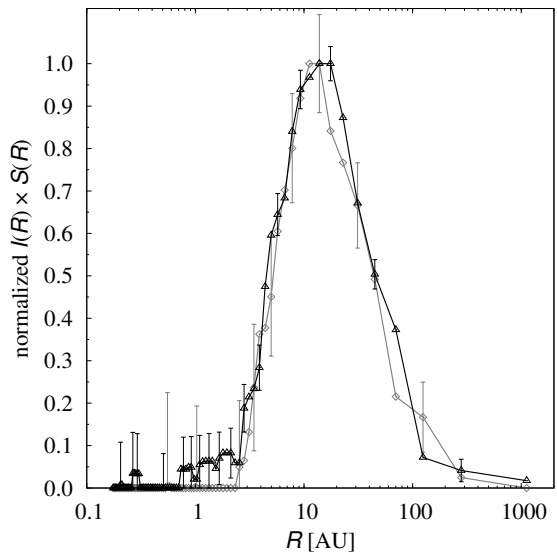}
\caption{The intensity versus radius graphs $I(R)$ derived from the line profiles
of the [\ion{O}{1}] 6300 (black, triangles) and 6363~\r{A} (gray, diamonds) emission lines.
Intensities have been multiplied by surface $S(R)$ of a ring of the disk at distance $R$
and normalized to peak intensity.
Error bars, derived from the S/N of the spectra, are shown every third data point.
A distinct inner and outer disk component can clearly be seen in the data
derived from the [\ion{O}{1}] 6300~\r{A} line.
The boundary of these components corresponds to the estimated dust sublimation radius.
The inner component is not seen in the weaker 6363~\r{A} line.
\label{fig:ivsr}}
\end{figure}

We compute radial surface brightness profiles from the [\ion{O}{1}] lines at 6300 and 6363~\r{A},
adopting a stellar mass of $6.6~M_{\odot}$ and disk inclination of $50\degr$ (see Figure \ref{fig:ivsr}).
The most striking result is the clear dichotomy in the profile derived from the 6300~\r{A} line.
It features a shallow plateau from roughly 1 to 2.5~AU (corresponding to the high velocity wings of the emission line),
and a much stronger component from about 2.5~AU outwards, peaking around 20~AU (corresponding to the core of the line).
The dependence of these values on disk inclination is relatively weak,
with the division between the two components varying between 2 and 3~AU
for inclinations between $40\degr$ and $60\degr$ respectively.
The profile derived from the 6363~\r{A} line has a very similar behavior, as expected,
but does not show an inner component.
This can be explained by the intrinsic weakness of the 6363~\r{A} line
(the ratio of the Einstein transition rates of the lines is $A_{6300}/A_{6363}$ = 3.0),
and the slightly lower S/N achieved in this part of the spectrum,
which would blend any possible high velocity wings into the noise of the continuum.

The observed distribution of emitting atomic oxygen as a function of radius can be explained by a disk model
consisting of a weakly emitting inner disk and a strongly emitting outer disk.
The transition between these two components lies between approximately 2 and 3~AU.
The outer radius of the emission region cannot be derived accurately from the line profile,
since Keplerian rotational velocity goes to zero at large distances from the star.
Nonetheless, [\ion{O}{1}] emission is present up to at least 100~AU from the star.
At this distance, our spectral resolution is still sufficient to resolve the disk kinematics.

\section{Discussion and conclusions} \label{sect:discuss}

We have presented optical spectra of MWC~147, featuring various double-peaked emission lines.
The observed differences in line width are consistent with a highly stratified circumstellar structure
in which line shape is dominated by rotation. The double-peaked emission lines are highly symmetric 
and centered on the stellar radial velocity derived from \ion{He}{1} absorption lines.
These properties indicate emission from the surface of a rotating, circumstellar disk.
The observation of \ion{Mg}{2} emission close to the corotation radius
shows that this disk extends to very close to the stellar surface.

According to \citet{ack05}, [\ion{O}{1}] emission in HAeBe stars cannot be due to thermal excitation.
They suggest that photodissociation of OH molecules by UV radiation from the star
provides the excited neutral oxygen atoms.
The 6300~\r{A} to 5577~\r{A} line flux ratio of about 10 observed in MWC~147 is in agreement with
values predicted by \citet{sto00} for this nonthermal excitation mechanism in photodissociation regions.
However, we cannot exclude a thermal contribution to the lines for MWC~147,
which has a much higher luminosity than most of the stars discussed by \citet{ack05}.

Analysis of the [\ion{O}{1}] 6300~\r{A} line shows that the disk is composed of
a weakly emitting inner component and a more strongly emitting outer part,
with the break between the two components at 2 to 3~AU.
This matches the dust sublimation radius estimate of 2.7~AU
by \citet[][assuming a dust sublimation temperature of 1500~K]{kra07, kra08},
in the framework of their disk model composed of a flat inner gaseous disk and a flaring outer dust disk.
The change in disk geometry at the dust sublimation radius is likely to cause a steep density gradient
in the upper disk atmosphere, due to the strong change in disk scale height,
related to the decreased opacity of the inner disk.
If the hydrogen density rises above the critical value of $6 \times 10^9$~cm$^{-3}$ \citep{sto00},
collisional de-excitation hampers the radiative channel,
resulting in a (much) weaker [\ion{O}{1}] emission.
Densities around this critical value are plausible for the inner regions of protoplanetary disks
surrounding intermediate-mass pre-main sequence stars (van der Plas et al., in preparation, based on \citealt{woi09}).
Detailed modeling is needed to establish this connection quantitatively.
We note that the permitted \ion{O}{1} 8446~\r{A} line is much broader than the forbidden lines (Figure \ref{fig:lines})
and does not show a break at the velocity corresponding to the estimated dust sublimation radius.
This excludes a drastic change in the \ion{O}{1} abundance
as an explanation for the observed two-component [\ion{O}{1}] 6300~\r{A} line profile.
Furthermore, in a larger sample of HAe stars
the inner radius of the [\ion{O}{1}] emitting region derived from the [\ion{O}{1}] line shape
is similar to that of the dust sublimation radius (van der Plas, private communication),
as is the case for the strong component of the [\ion{O}{1}] emission in MWC~147.
Whether or not the inner disk in MWC~147 is an accretion disk as \citet{kra08} suggest is unclear.

It is interesting to compare the geometry of the disk of MWC~147 to those of less luminous HAeBe stars.
There is growing evidence from interferometric studies that the lower mass HAe stars have material inside the dust sublimation radius 
\citep[e.g.][]{ben10} whose nature is not well constrained.
Both gas and refractory dust have been proposed.
The higher luminosity of MWC~147 forces the dust sublimation radius to larger distance,
allowing the inner regions to appear more pronounced in imaging and spectroscopy.
The case of MWC~147 clearly proves that an innermost gaseous disk is possible.

Analysis of the infrared SED of MWC~147 by \citet{ver10}
suggests either a self-shadowed disk or a disk with a small but not well constrained outer radius.
These authors show that emission from Polycyclic Aromatic Hydrocarbons (PAHs) detected in a Spitzer IRS spectrum
arises from the environment and is weak or absent in the disk.
This implies that MWC~147 does not follow the trend that strong emission in [\ion{O}{1}] 6300~\r{A}
is usually seen in disks with strong PAH emission \citep{ack05}.
Strong emission in these gas tracers is usually interpreted as evidence for a flaring disk.
If the presence of [\ion{O}{1}] 6300~\r{A} emission up to distances of $\sim100$~AU is due to
non-thermally excited \ion{O}{1}, then a gaseous flaring disk must be present extending to these distances.
This would favor a scenario in which PAHs are destroyed by the strong UV radiation field of the star.

The age of the disk \citep[$3.2 \times 10^5$~yr,][]{her04} is comparable to
the expected evaporation timescale for a $6 - 7~M_{\sun}$ star \citep[$\sim10^5$~yr,][]{gor09}.
It is therefore expected to be in a phase of significant evaporation.
Our observations constrain the disk to extend to at least 100~AU,
suggesting that at this distance it has not yet significantly been eroded by the stellar radiation field.

\acknowledgments

We would like to thank the anonymous referee for a careful review of the letter.
This study is based on observations made with the Mercator Telescope,
operated on the island of La Palma by the Flemish Community, at the
Spanish Observatorio del Roque de los Muchachos of the Instituto de
Astrof\'{i}sica de Canarias.

The HERMES project is a collaboration of the K.U.Leuven, ULB and ROB of Belgium
with smaller contributions from the Tautenburg Landessterwarte and the Geneva
observatory. The Hermes team acknowledges support from the Fund for Scientific
Research of Flanders (FWO), from the Research Council of K.U.Leuven, from the
Fonds National Recherches Scientific and from the Royal Observatory of Belgium.

{\it Facilities:} \facility{Mercator1.2m (HERMES)}.

\end{document}